\documentclass{elsart}

\usepackage{graphicx}
\begin{document}

\begin{frontmatter}

\title{Observational Evidence for Two Cosmological Predictions
Made by Bit-String Physics}
\thanks{Work supported by Department of Energy contract DE--AC03--76SF00515.}

\author{H. Pierre Noyes}

\address{Stanford Linear Accelerator Center\\ Stanford
University, Stanford, CA 94309}

\begin{abstract}
A decade ago bit-string physics predicted that the baryon/photon
ratio at the time of nucleogenesis $\eta= 1/256^4$ and that the
dark matter/baryonic matter ratio $\Omega_{DM}/\Omega_B= 12.7$.
Accepting that the normalized Hubble constant is constrained
observationally to lie in the range $0.6 < h_0 < 0.8$, this
translates into a prediction that $0.325 > \Omega_M > 0.183 $.
This and a prediction by E.D.Jones, using a model-independent argument
and ideas with which bit-string physics is not inconsistent, that the cosmological constant $\Omega_{\Lambda}=0.6\pm 0.1$ are in reasonable agreement with recent cosmological observations, including the BOOMERANG data.
\end{abstract}


\end{frontmatter}

%\begin{center}
%\emph{Submitted to New Astronomy}
%\end{center}

\emph{Bit-string physics} (BSP) is an alternative approach to
natural philosophy based on four principles:

1. Happenings can be distinguished from nothing.

2. Happenings are the same or different.

3. Happenings can be recorded and these records can be
re-examined.

4. In the absence of further information, all happenings are
equally probable.

This research program has a long history \cite{Noyes97a}, starting
with the discovery of the \emph{combinatorial hierarchy} by
A.F. Parker-Rhodes in 1961 \cite{Parker-Rhodes62}, but has not
attracted much attention in the mainstream literature. One
difficulty, according to several of our critics, is that although
BSP has produced approximate values for well known physical
parameters it has not led to quantitative predictions prior to
observation. Here we meet this difficulty in an unexpected manner
by showing that recent cosmological observations support two
predictions made about a decade ago when there was no available
way to compare them with observation.

I now use the following arguments to develop a growing universe of
bit-strings from our basic principles. If nothing happens and we
have no structure to generate happenings, nothing will happen. So
we start with a happening. But we can't yet know whether the
``nothing happening'' we are, by hypothesis, able to distinguish
from this first happening is ``elsewhere'', or ``elsewhen'',
or both, or\dots . I assume that we should use the simplest
possible mathematical structures to model and develop our basic
concepts. Distinguishing a happening from nothing happening is
simply modeled by bit multiplication, i.e. $0\cdot 0= 1\cdot 0
=0\cdot 1=0$; $1\cdot 1 = 1$. This implies that the two cases can
be compared, so we know that our ``start'' is most simply
represented by a ``1'' \emph{and} a ``0''. We record the two
symbols (Principle 3), but still don't know which symbol refers to
a happening and which to nothing happening. We now model the
comparison (Principle 2) by bit addition $1 \oplus 1 =0 =0 \oplus
0$; $1 \oplus 0 =1=0\oplus 1$ (i.e. addition modulo 2, XOR,
symmetric difference,... or, as it is referred to in the ANPA
research program, \emph{discrimination}). We know if we compare
the two symbols already recorded using this operation we will
necessarily produce a ``1'' which we record.  How we keep track of
what has happened should not affect the subsequent development, so
we record the two ``1'' 's and the ``0'' as a column, but ascribe
no significance to the order of the three symbols in the column.

We now introduce a simple algorithm called \emph{program universe}
\cite{Manthey86,Noyes&McGoveran89,Noyes97b} which generates a
growing universe of bit-strings from the starting column. This
algorithm will allow us to make arbitrary choices of either a
symbol or of a row in the table, based on Principle 4. In other
words, the algorithm contains a ``random number generator''. Of
course for any realized computer simulation this can only be
``pseudo-random''. At each step this universe contains $P(S)$
strings of length $S$. As argued above the starting point required
by our principles is three rows of length one, i.e. $P(1)=3$,
containing two ``1'' 's and one ``0''. The algorithm is very
simple, as can be seen from the flow diagram in Fig. 1.
\begin{figure}
\centering
\includegraphics{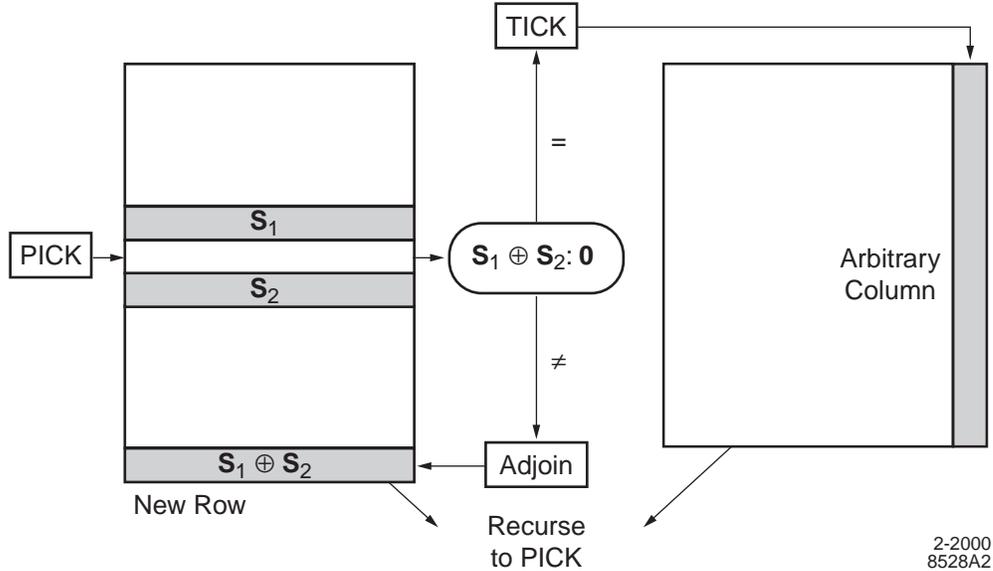}

\caption{Flow diagram for constructing a bit-string universe
growing by one row or one column at a time (see text).}
\label{fig1}
\end{figure}
We start with a rectangular block of rows and columns containing
only the bits ``0'' and ``1''. We then pick two rows arbitrarily
and if their discriminant is non-null, adjoin it to the table as a
new row and recurse to picking two arbitrary rows (PICK). If the
discriminant is null, we simply adjoin an arbitrary column
(Bernoulli sequence) to the table and recurse to PICK. That this
model contains arbitrary elements and (if interpretable in terms
of known aspects of the practice of physics) an historical record
ordered by the number of TICK's, or equivalently by the row
length, should be clear from the outset. The forging of rules that
will indeed connect the model to the \emph{actual} practice of
physics is the primary research problem that has engaged me ever
since the model was created.

Program universe provides a separation into a conserved set of
``labels'', and a growing set of ``contents'' which can be thought
of as the space-time ``addresses'' to which these labels refer. To
see this, think of all the left-hand, finite length $S$ portions
of the strings which exist when the program TICKs and the
string-length goes from $S$ to $S+1$. Call these \emph{labels} of
length $L=S$, and the number of them at the critical TICK
$P_0(L)$, and the string length before the tick $S_L$. Further
PICKs and TICKs can only add to this set of labels of this fixed
length those which can be produced from it by pairwise
discrimination, with no impact from the (growing in length and
number) set of content labels; the length of the content (address)
part of the string is  $S_C=S-L > 0$. If $N_I \leq P_0(S_L)$ of
these labels are \emph{discriminately independent}, then the
maximum number of distinct labels they can generate, no matter how
long program universe runs, will be $2^{N_I} - 1$, because this is
the maximum number of ways we can choose combinations of $N_I$
distinct things taking them $1,2,\ldots,N_I$ at a time. We will
interpret this fixed number of allowable labels as a
representation of the  quantum numbers of systems of
 ``elementary particles'' present in our bit-string universe
and use the growing content-strings to represent their (finite and
discrete) locations in an expanding space-time description of the
universe.

\begin{figure}
\centering
\includegraphics{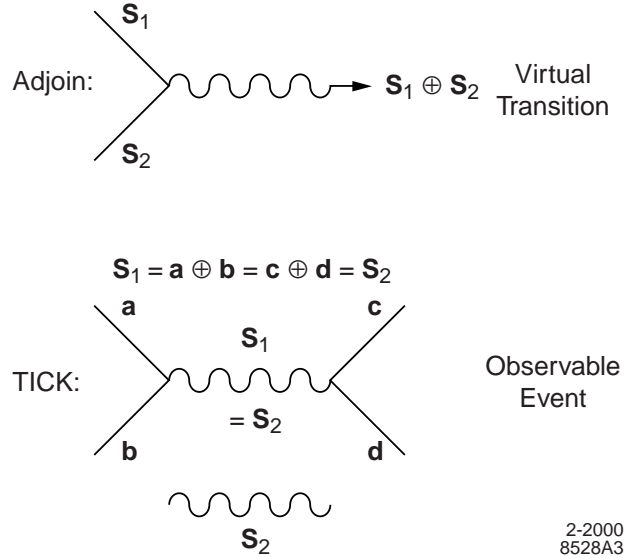}

\caption{Interpretation of the Adjoin and TICK operations of
Program Universe as Feynman diagrams(see text).} \label{fig2}
\end{figure}

This label-content schema then allows us to interpret the events
which lead to  TICK as four-leg Feynman diagrams representing a
stationary state scattering process. Note that for us to find out
that the two strings found by PICK are the same, we must either
pick the same string twice or at some previous step have produced
(by discrimination) and adjoined the string which is now the same
as the second one picked. Short-circuiting and reordering the
actual route by which my current interpretation of this model was
arrived at, we note that the two basic operations in the model
which provide locally novel bit-strings (Adjoin and TICK) are
isomorphic, respectively, to a three-leg or a four-leg Feynman
diagram. This is illustrated in Fig. \ref{fig2}. Note that the internal
(exchanged particle)
state in the four-leg Feynman diagram is \emph{necessarily}
accompanied by an identical (but distinct) ``spectator'' somewhere
else in the (coherent) memory.

Although the paper is not presented in bit-string language, a
little thought about the solution of a relativistic three body
scattering problem Ed Jones and I have found
\cite{Noyes&Jones2000} shows that the driving term ($>-<\atop -$)
is always a four-leg Feynman diagram ($>-<$) plus a spectator ($\
- \ $) whose quantum numbers are \emph{identical} with the quantum
numbers of the particle in the intermediate state connecting the
two vertices. We are particularly pleased that the observable
events created by program universe turn out to provide \emph{two}
locally identical but distinct strings (states) needed as the
starting point for this scattering theory \cite{Noyes&Jones2000}.
We do not have space here to explain how, in the more detailed
dynamical interpretation, the three-leg diagrams conserve
(relativistic) 3-momentum but not necessarily energy  (like vacuum
fluctuations) while the four-leg diagrams conserve both 3-momentum
and energy and hence are candidates for potentially observable
events. But we \emph{do} need to explain how this interpretation
of program universe does connect up with the work on the
combinatorial hierarchy.

\newpage

At this point we need a guiding principle to show us how we can
``chunk'' the growing information content provided by the
discriminate closure of the label portion of the strings in such a
way as to generate a hierarchical representation of the quantum
numbers that these labels represent. Following a suggestion of
David McGoveran's \cite{McGoveran98}, we note that \emph{we can
guarantee that the representation has a coordinate basis and
supports linear operators by mapping it to square matrices}.

The mapping scheme originally used by Amson, Bastin, Kilmister and
Parker-Rhodes \cite{Bastin66} satisfies this requirement. This
scheme also uses bit-string discrimination, multiplication (hence
the \emph{field} $Z_2$), and \emph{discriminate closure}; these
are the basic formal elements we ``derived'' above from our basic
principles. First note, as mentioned above, that any set of $n$
discriminately independent (\emph{d.i.}) strings will generate
exactly $2^n -1$ discriminately closed subsets (\emph{dcss}).
Start with two d.i. strings $\mathbf{a}$, $\mathbf{b}$. These generate
three d.i. subsets, namely $\{\mathbf{a} \}$, $\{\mathbf{b} \}$, $\{
\mathbf{a}, \mathbf{b}, \mathbf{a}\oplus \mathbf{b} \}$. Require each dcss ($\{
\  \}$) to contain only the eigenvector(s), of three $2\times 2$
\emph{mapping matrices} which (1) are non-singular (do not map
onto zero) and (2) are d.i. Rearrange these as strings. They will
then generate seven dcss. Map these by seven d.i. $4\times 4$
matrices, which meet the same criteria (1) and (2) just given.
Rearrange these as  seven d.i. strings of length 16. These
generate $127=2^7-1$ dcss. These can be mapped by 127 $16\times
16$ d.i. mapping matrices, which, rearranged as strings of length
256, generate $2^{127}-1 \approx 1.7\times 10^{38}$ dcss. But
these cannot be mapped by $256\times 256$ d.i. matrices because
there are at most $256^2$ such matrices and $256^2 \ll 2^{127}-1$.
Thus this \emph{combinatorial hierarchy} terminates at the fourth
level. The mapping matrices are not unique, but exist, as has been
proved by direct construction and an abstract proof
\cite{Bastinetal79}. It is easy to see that the four level
hierarchy constructed by these rules is \emph{unique} because
starting with d.i. strings of length 3 or 4 generates only two
levels and the dcss generated by starting with d.i. strings of
length 5 or greater cannot be mapped.

Making physical sense out of these numbers is a long story
\cite{Noyes97a}. In order to underpin our claim that we can model
a finite particle number version of relativistic quantum mechanics
with particle creation, etc. using bit-strings we give on the next
page the predictions of coupling constants and mass ratios
calculated using our theory. As in any mass, length, time theory
we are allowed three empirical, dimensional constants which are
measured by standard techniques to connect our abstract theory to
measurement. These we take to be the mass of the proton $m_p$,
Planck's constant $\hbar$ and the velocity of light $c$.
Everything else is calculated. Agreement with observation, given
below, is not perfect; we believe it is impressive. For more
detail see \cite{Noyes97a}.

\section*{Comparison of Bit-String Predictions of Coupling
Constants and Mass Ratios with Experiment}

$$G_N^{-1} {\hbar c \over m_p^2}= [2^{127} + 136]\times \left[\mathbf{
1-{1\over 3\cdot 7\cdot 10}}\right] =1.693 \ 31\ldots\times 10^{38}$$

$$ \mbox{experiment} =1.693 \ \mathbf{58}(21) \times 10^{38}$$

$$\alpha^{-1} (m_e)= 137\times \left[\mathbf{1- {1 \over 30 \times 127}}\right]^{-1} =
137.0359 \ \mathbf{674} \ldots. $$

$$  \mbox{experiment} =137.0359\ 895(61)$$

$$G_Fm_p^2/\hbar c = [256^2\sqrt{2}]^{-1}\times \left[\mathbf{[1 - {1 \over
3\cdot 7}}\right] =1.02 \ \mathbf{758}\ldots\times 10^{-5}$$

$$ \mbox{experiment} = 1.02 \ 682(2)\times 10^{-5}$$

$$sin^2\theta _\mathrm{Weak}= 0.25 \left[\mathbf{1- {1 \over 3\cdot 7}}\right]^2
 = 0.2267\ldots$$

$$\mbox{experiment} =0.22 \mathbf{59}(46)$$

\begin{equation}
{m_p\over m_e} ={137 \pi\over \langle x(1-x)\rangle \left\langle{1\over y}
\right\rangle}= {137 \pi\over
\left({3\over 14}\right) \left[1 + {2\over 7} + {4\over 49}\right] \left({4\over 5}\right)}
 = 1836.15 \ \mathbf{1497}\ldots
\end{equation}

$$ \mbox{experiment} =1836.15\ 2701(37)$$

$$m_{\pi}^{\pm}/ m_e =275\left[\mathbf{1 - {2\over 2\cdot 3 \cdot 7\cdot 7}}\right]
= 273.12 \ \mathbf{92}\ldots$$

$$ \mbox{experiment} = 273.12 \ 67(4)$$

\newpage

$$m_{\pi ^0}/m_e =274 \left[\mathbf{1- {3\over 2\cdot 3 \cdot 7 \cdot 2}}\right]
 =264.2 \ \mathbf{143\ldots}$$

$$\mbox{experiment} =264.1 \ 373(6)]$$

$$m_{\mu}/m_e =3\cdot 7\cdot 10 \left[1-{3\over 3\cdot 7\cdot 10}\right]= 207$$

$$ \mbox{experiment} =206.768 \ 26(13)$$

$$G^2_{\pi N\bar N}= \left[\left({2 M_N\over
m_{\pi}}\right)^2-1\right]^{{1\over 2}}=[195]^{{1\over 2}}=13.96\ldots .$$

$$\mbox{experiment} = 13.3(3), \mbox{ or  greater   than }\ 13.9 $$

\begin{figure}
\centering
\includegraphics{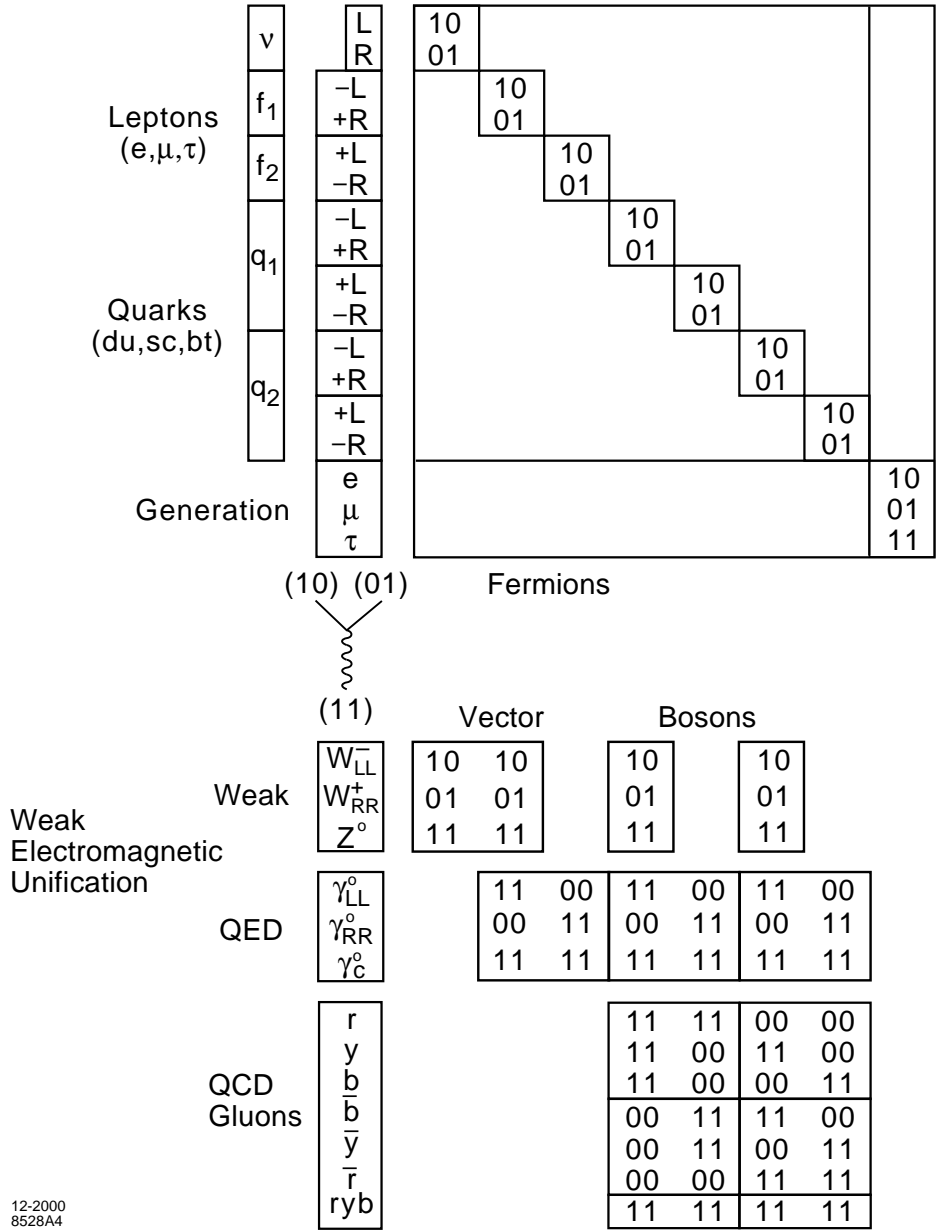}

\caption{Skeleton of a label scheme for labels of length 16
which conveys the same quantum number information as the standard
model of quarks and leptons.} \label{fig7}
\end{figure}

\noindent Making the case that these constructions give us the
quantum numbers of the standard model of quarks and leptons with
exactly 3 generations has only been sketched \cite{Noyes94}.A
tentative bit-string representation of the quantum numbers of the
(three generation) standard model of quarks and leptons using
bit-string labels of length sixteen is given in Fig. \ref{fig7}.

Fortunately we do not require the completely worked out scheme to
make interesting cosmological predictions.  The ratio of dark to
``visible''  (i.e. electromagnetically interacting) matter is the
easiest to see. The electromagnetic interaction first comes in
when we have constructed the first three levels giving 3+7+127
=137 dcss, one of which is identified with electromagnetic
interactions because it occurs with probability $1/137 \approx
e^2/\hbar c$. But the construction must first complete the first
two levels giving $3+7=10$ dcss. Since the construction is
``random'' and this will happen many, many times as program
universe grinds along, we will get the 10 non-electromagnetically
interacting labels 127/10 times as often as we get the
electromagnetically interacting labels. Our prediction of
$M_{DM}/M_{B}= 12.7$ is that naive.

The $1/256^4$ prediction for $N_{B}/N_{\gamma}$ is comparably
naive. Our partially worked out scheme of relating bit-string
events to particle physics \cite{Noyes94,Noyes97a}, makes it clear
that photons, both as labels (which communicate with
particle-antiparticle pairs) and as content strings will contain
equal numbers of zeros and ones in appropriately specified
portions of the strings. Consequently they can be readily
identified as the most probable entities in any assemblage of
strings generated by whatever pseudo-random number generator is
used to construct the arbitrary actions and bit-strings needed in
actually running program universe. This scheme also makes the
simplest representation of fermions and anti-fermions contain one
more ``1'' and one less ``0'' than the photons (or \emph{visa
versa}). (Which we call ``fermions'' and which ``anti-fermions''
is, to begin with, an arbitrary choice of nomenclature.) Since our
dynamics insures conventional quantum number conservation by
construction, the problem is to show how program universe
introduces a bias between ``0'' 's and ``1'' 's once the full
interaction scheme is developed; this problem is analogous to the
corresponding problem in conventional theories. However, our
theory requires no ``fine tuning''.

We saw above that program universe has to start out with two one's
and one zero. Subsequent PICKs and TICKs are sufficiently
``random'' to insure that (at least statistically) we will
generate an equal number of zeros and ones, apart from the initial
bias giving an extra one. Once the label length of 256 is reached,
which is the label length required by the combinatorial hierarchy
mapping scheme discussed above, and sufficient space-time
structure (``content strings'') generated and interacted to
achieve thermal equilibrium, this label bias for a 1 compared to
equal numbers of zeros and ones will persist for 1 in 256 labels.
We must now count the number of equilibrium processes leading to
baryon-photon scattering relative to the number leading to
photon-photon scattering.
\begin{figure}
\centering
\includegraphics{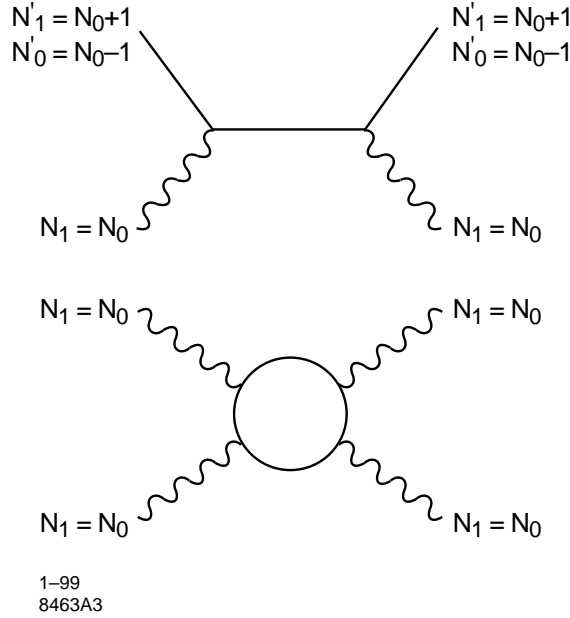}

\caption{Comparison of bit-string labeled processes after the
label length is fixed at 256 interpreted as baryon ($N_1'=N_0+1$)
photon ($N_1=N_0$) and photon-photon scattering. Here $N_1$ and
$N_0$ symbolize, respectively,  the number of ones and zeros in
the label part of the string  (which is of length 256). Program
universe guarantees that, in the absence of further
considerations, the content part of the strings will have an equal
number of zeros and ones with very high probability as the string
length (universe) grows. } \label{fig3}
\end{figure}
We start from the most probable and the next most probable classes
of scattering processes, which are presented in Figure \ref{fig3}. Because
baryon number is conserved, and the even-odd character of the
labels is conserved by discrimination, we interpret the bias as
specifying baryon number for one of the 256 labels in the initial
(or equivalently the final) state. This requires the baryon bias
of 1 to appear in one and only one of the four initial (or final,
because of baryon number conservation) state labels of length 256.
Then, simple case counting gives the baryon to photon ratio as
$1/256^{4}$ for any representational scheme which requires photons
to have equal numbers of zero's and ones, and assigns baryon
number to any one of the 256 slots in the label. Of course this
conclusion rests on the interpretation of the strings causing
observable TICK's as four-leg Feynman diagrams; that
interpretation still needs some work on the details. As a trivial
example of how this could work for labels of much shorter length
take the baryon-antibaryon-photon vertex to be $\mathbf{B}\oplus \mathbf{
\bar{B}} \oplus \mathbf{\gamma} = \mathbf{0}$ with $\mathbf{B}=(1110)$,
$\mathbf{\bar{B}}=(0010)$ and $\mathbf{\gamma}= (1100)$. We conclude
that, in the absence of further information, $1/256^4$ is the
program universe prediction for the baryon-photon ratio at the
time of big bang nucleosynthesis.

\begin{figure}
\centering
\includegraphics{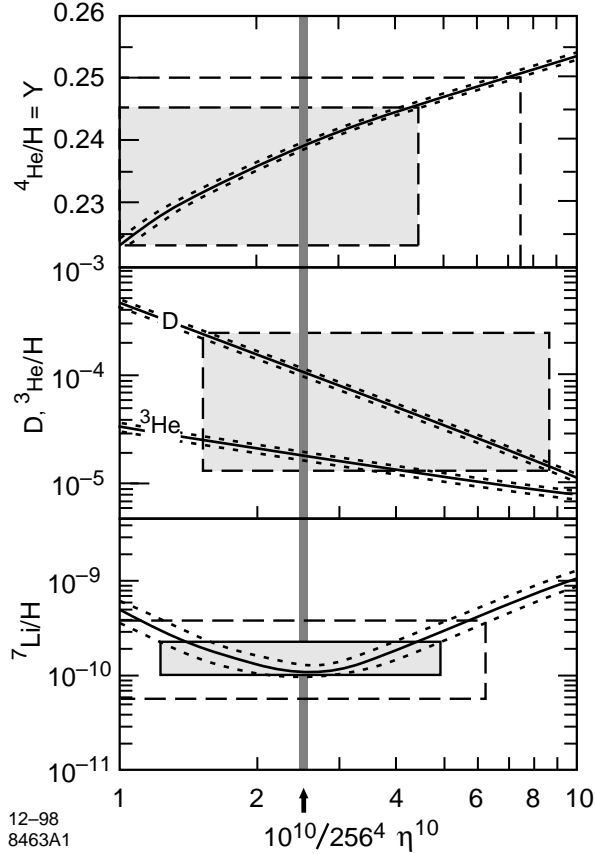}

\caption{Comparison of the bit-string physics prediction that
$\eta= 256^{-4}$ with accepted limits on the cosmic abundances as
given by Olive and Schramm in  \cite{PDG98}, p.~119.} \label{fig4}
\end{figure}

We now demonstrate that our two predictions can be compared with
observation, by showing that, together with the currently accepted
value of the Hubble constant, they allow us to predict the
normalized matter density $\Omega_M$ with reasonable precision. We
recall that the predictions are that (a) the ratio of baryons to
photons was $\eta= 1/256^4 =2.328\ldots\times
10^{-10}=10^{-10}\eta^{10}$ at the time of nucleogenesis and that
(b) $\Omega_{DM}/\Omega_B=127/10=12.7$. Comparison of prediction
(a) with observation is straightforward, as is illustrated in
Figure \ref{fig4}.

Comparison with observation of prediction (b) that the ratio of
dark to baryonic matter is \emph{not} straightforward, as was
clear at DM98. This question remained unresolved at DM2000.
However, according to the standard cosmological model, the
baryon-photon ratio remains fixed \emph{after} nucleogenesis. In
the theory I am relying on, the same is true of the of the dark
matter to baryon ratio. Consequently, \emph{if} we know the Hubble
constant, \emph{and} assume that only dark and baryonic matter
contribute, the normalized matter parameter $\Omega_M$ can \emph{
also} be predicted, as we now demonstrate.

We know from the currently observed photon density (calculated
from the observed $2.728\,^\circ$K cosmic background radiation) that
the normalized baryon density is given by \cite{Olive98}
\begin{equation}
\Omega_B = 3.67\times 10^{-3}\eta_{10}h_0^{-2}
\end{equation}
and hence, from our prediction and assumptions about dark matter,
that the total mass density will be 13.7 times as large. Therefore
we have that
\begin{equation}
\Omega_M= 0.117 h_0^{-2} \ .
\end{equation}
Hence, for $0.8 \geq h_0 \geq 0.6$  \cite{Hogan98}, $\Omega_M$
runs from $0.183$ to $0.325$. This clearly puts no restriction on
$\Omega_{\Lambda}$.

Our second constraint comes from integrating the scaled
Friedman-Robertson-Walker (FRW) equations from a time after the
expansion becomes matter dominated with no pressure to the
present. Here we assume that this initial time is close enough to
zero on the time scale of the integration so that the lower limit
of integration can be approximated by zero \cite{Primackinp}. Then
the age of the universe as a function of the current values of
$\Omega_M$ and $\Omega_{\Lambda}$ is given by
\begin{eqnarray}
t_0&=&9.78 h_0^{-1}f(\Omega_M,\Omega_{\Lambda}) \ Gyr\nonumber\\
&=&9.78 h_0^{-1}f(0.117h_0^{-2},\Omega_{\Lambda}) \ \mbox{Gyr}
\end{eqnarray}
where
\begin{equation}
f(\Omega_M,\Omega_{\Lambda})=\int_0^1dx \sqrt{{x\over \Omega_M
+(1-\Omega_M-\Omega_{\Lambda})x +\Omega_{\Lambda}x^3}} .
\end{equation}
For the two limiting values of $h_0$, we see that
\begin{eqnarray}
h_0&=&0.8,\ \ t_0=12.2f(0.183,\Omega_{\Lambda}) \ \mbox{Gyr}
\nonumber\\
h_0&=&0.6,\ \ t_0=16.3f(0.325,\Omega_{\Lambda}) \ \mbox{Gyr}  .
\end{eqnarray}

The results are plotted in Figure 6 in comparison with data
available in 1998. We emphasize that these predictions were made
and published over a decade ago when the observational data were
vague and the theoretical climate of opinion was very different
from what it is now. The figure reproduced here was presented at
ANPA20 (Sept. 3--8, 1998, Cambridge, England) and given wider
circulation in \cite{Noyes99}. The calculation that
$\Omega_{\Lambda}=0.6$ included in the figure and briefly
discussed below was made by E.D.~Jones before there was any
observational evidence for a cosmological constant, let alone a
positive one \cite{Jones97}.

\begin{figure}
\centering
\includegraphics{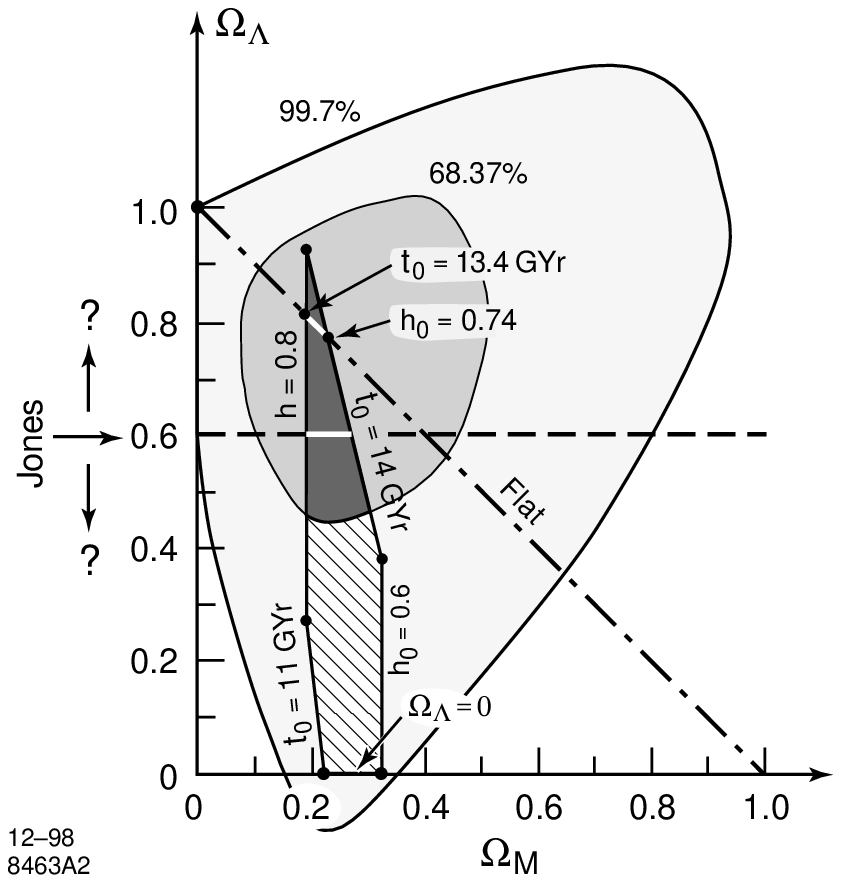}

\caption{Limits on $(\Omega_M, \Omega_{\Lambda})$  set by
combining the Supernovae Type Ia data from Perlmutter, et al. with
the Cosmic Ray Background Experiment (COBE) satellite data as
quoted by Glanz
 \cite{Glanz98a} (dotted curves at the 68.37\% and 99.7\% confidence
levels) compared with the predictions of bit-string physics that
$\eta_{10} = 10^{10}/256^4$ (cf. Fig. 5) and $\Omega_\mathrm{Dark}/
\Omega_B =12.7$. We accept the constraints on the scaled
Hubble constant $h_0=0.7\pm 0.1$  \cite{Groometal98} and on the
age of the universe $t_0 = 12.5\pm1.5 \ \mbox{Gyr}$ (solid lines). We
include the predicted constraint $\Omega_{\Lambda} > 0)$. The
Jones estimate of $\Omega_{\Lambda}= 0.6$ is indicated, but the
uncertainty was not available in 1998.} \label{fig5}
\end{figure}

The precision of the relevant observational results had improved
considerably by DM2000,where the preliminary version of this paper
was presented \cite{Noyes00}. Using an analysis due to Lineweaver
\cite{Lineweaver99}, our prediction was still in excellent
agreement with observation. Thanks to still more recent analyses
of the BOOMERANG data and other observations \cite{PDG00}, we now
face a still more stringent test, as shown in Figure 7.

\begin{figure}
\centering
\includegraphics{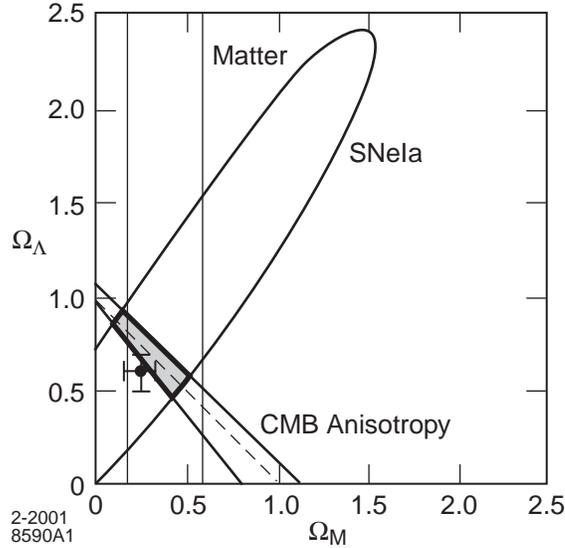}

\caption{Comparison of the bit-string physics and the Jones
predictions with currently accepted cosmological data.}
\label{fig6}
\end{figure}

As in 1998, we find it useful to make use of an unpublished
prediction of $\Omega_{\Lambda}$ by E.D.~Jones \cite{Jones97}.
In contrast to the situation then, he is now prepared to set definite limits on his prediction based on considerations of his calculation
made while preparing his paper for publication.  Further, the estimate given above ($\Omega_{\Lambda} \sim 0.6\pm 0.1$), which was made before and independent of our calculations reported above, falls squarely in the middle of the region allowed in 1998, and continues to do so despite the remarkable observational progress made in the interim.

The Jones calculation depends only on self-consistency arguments and requires the external input of an astrophysical quantity, such as the mass of the universe in Planck units, to calibrate the parameters. As a model-independent argument, it can not calculate the needed input from some first principles. Thus, the argument presumably can be improved by combining it with the bit-string physics model. Research in that direction is in progress.

We find it remarkable that the four epistemological principles
with which we start seem to lead to (a) the overall structure of
elementary particle physics together with a number of the basic
constants calculated using no free parameters and (b) the gross
cosmological state of the universe as measured by $\Omega_M$ and
$\Omega_{\Lambda}$. We hope that these facts will motivate others
to investigate how this might come about.


\begin{thebibliography}{99}

\bibitem{Bastin66}
T. Bastin, ``On the Scale Constants of Physics'', \emph{Studia
Philosophica Gandensia}, \textbf{4} (1966), 77.

\bibitem{Bastinetal79}
T. Bastin, H.P. Noyes, J. Amson and C.W. Kilmister, \emph{Int'l.
J. Theor. Phys.}, \textbf{18} (1979), 445--488.

\bibitem{Glanz98a}
J. Glanz, \emph{Science}, \textbf{280} 15 May 1998, 1008.

\bibitem{Glanz98b}
J. Glanz, \emph{Science}, \textbf{282} 13 September 1998, 1247.

\bibitem{Groometal98}
D.E. Groom, et al, ``Astrophysical Constants'', in \cite{PDG98},
p. 70.

\bibitem{Hogan98}
C.J. Hogan, ``The Hubble Constant'', in \cite{PDG98}, pp 122--124.

\bibitem{Jones97}
E.D. Jones, private communications to HPN starting in 1997; this
fact, and a seminar on the type Ia supernovae data by Goldhaber at
SLAC were primary motivations for HPN to attend DM98 and make the
presentation already cited \cite{Noyes99} that summer at ANPA20
(Sept. 1998).

\bibitem{Lineweaver99}
C.H. Lineweaver, \emph{Science}, \textbf{284} (1999), 1503--1507.

\bibitem{Manthey86}
M.J. Manthey, ``Program Universe'' in  SLAC-PUB-4008, June 1986
(Part of Proc. ANPA 7), pp 101--110.


\bibitem{McGoveranBSP}
D.O. McGoveran, in \emph{Proc. ANPA 10,11}, reprinted in
\cite{Noyesinp}, Ch. 6.

\bibitem{McGoveran98}
D.O. McGoveran, private communication to HPN November 19, 1998.
McGoveran has tried to get this point across to this author, and
to others committed to the ANPA research program, for several
years.

\bibitem{McGoveran&Noyes91}
D.O. McGoveran and H.P. Noyes, \emph{Physics Essays}, \textbf{4} (1991),
115--120.

\bibitem{Noyes94}
H.P. Noyes, ``Bit-String Physics, a Novel `Theory of Everything'$\,$'',
in \emph{Proc. Workshop on Physics and Computation (PhysComp
'94)}, D. Matzke, ed., 94, Los Amitos, CA: IEEE Computer Society
Press, 1994, pp. 88--94 and SLAC-PUB-6509. Aug 1994.


\bibitem{Noyes97a}
H.P. Noyes, ``A Short Introduction to BIT-STRING PHYSICS'', in \emph{
Merologies, Proc. ANPA 18}, T.L.Etter, ed; [available from ANPA
c/o K. Bowden, Theoretical Physics Research Unit, Birkbeck College,
Malet St., London WC1E 7HX]; SLAC-PUB-7205, (June 1997) and
hep-th-970702. This reference contains reasonably complete citations of
the earlier literature.

\bibitem{Noyes97b}
See \cite{Noyes97a}, Sect. 2, pp 28--32, for a brief discussion of
\emph{program universe} and a guide to the literature.

\bibitem{Noyes99}
H. P. Noyes, ``Program Universe and Recent Cosmological Results''
SLAC-PUB-8030 (Jan 99), [gr-qc/9901022]; it also appeared in the
Proceedings of the 20$^{th}$ annual meeting of the Alternative
Natural Philosophy Association, \emph{Aspects II}, K.G. Bowden,
Ed. pp 192--214 [available from ANPA c/o K.Bowden, Theoretical
Physics Research Unit, Birkbeck College, Malet St., London WC1E
7HX].

\bibitem{Noyes00}
H.P. Noyes, SLAC-PUB-8327(March 2000), SLAC-PUB-8342(April 2000)
and in \emph{Proc. DM2000}, Springer(in press).

\bibitem{Noyesinp}
H.P. Noyes, \emph{BIT-STRING PHYSICS: An Alternative Approach to
Natural Philosophy}, World Scientific, Singapore (in press).

\bibitem{Noyes&Jones2000}
H.P. Noyes and E.D. Jones, \emph{Few Body Systems}, \textbf{27} (1999),
123--139 and hep-th-971077.


\bibitem{Noyes&McGoveran89}
H.P.Noyes and D.O.McGoveran, \emph{Physics Essays}, \textbf{2} (1989), 76,
and SLAC-PUB-4528, Oct 1989.


\bibitem{Olive98}
K.A. Olive, ``Big-bang Cosmology'',  in \cite{PDG98}, pp 117--118.

\bibitem{Olive&Schramm98}
K.A. Olive and D.N. Schramm(dec.), ``Big-bang Nucleosynthesis'',
in \cite{PDG98}, pp 119--121.

\bibitem{Parker-Rhodes62}
A.F. Parker-Rhodes, ``Hierarchies of Descriptive Levels in Physical
Theory'', Cambridge Language Research Unit, internal document
I.S.U.7, Paper I, 15 January 1962; reprinted, together with
comments by John Amson in K. Bowden, ed. \emph{Int.J. General
Systems}, \textbf{27} Nos. 1-3(1998), pp 57--80.

\bibitem{PDG98}
Particle Data Group, ``Review of Particle Properties'', \emph{
 Phys.Euro. J.} \textbf{C 3} (1998), 1--794.

\bibitem{PDG00}
Particle Data Group, ``Review of Particle Properties'', \emph{
 Phys.Euro. J.} \textbf{C 15}, No. 1--4(2000), p127, Fig. 15.3.

\bibitem{Primackinp}
J.R. Primack, ``Dark Matter and Structure Formation'', in \emph{
Formation of Structure in the Universe}, Proc. of the Jerusalem
Winter School 1996, A. Dekel and J. P. Ostriker, eds. Cambridge
University Press (in Press) and astro-ph/9797285v2 25 Jul 1997,


\end{thebibliography}
\end{document}